\documentclass[final,5p,times,twocolumn]{elsarticle}
\usepackage{graphicx}
\usepackage[dvips,usenames]{color}
\usepackage{amsmath,amssymb}
\newcommand{\bs}{\begin{sloppypar}} \newcommand{\es}{\end{sloppypar}}
\def\beq{\begin{eqnarray}} \def\eeq{\end{eqnarray}}
\def\beqstar{\begin{eqnarray*}} \def\eeqstar{\end{eqnarray*}}
\newcommand{\bal}{\begin{align}}
\newcommand{\eal}{\end{align}}
\newcommand{\beqe}{\begin{equation}} \newcommand{\eeqe}{\end{equation}}
\newcommand{\p}[1]{(\ref{#1})}
\newcommand{\tbf}{\textbf}

\journal{Physics Letters B}

\begin{document}

\begin{frontmatter}



\title{Anisotropic pressure in dense neutron matter \\ under the presence of
a strong magnetic field}


\author[lab1,lab1a]{A. A.~Isayev\corref{cor1}}
\ead{isayev@kipt.kharkov.ua}
\author[lab2]{J.~Yang\corref{cor2}}
\ead{jyang@ewha.ac.kr}

\cortext[cor1]{Principal corresponding author}
\cortext[cor2]{Corresponding author}

\address[lab1]{Kharkov Institute of
Physics and Technology, Academicheskaya Street 1,
 Kharkov, 61108, Ukraine}
\address[lab1a]{Kharkov National University, Svobody Sq., 4, Kharkov, 61077,
Ukraine}
\address[lab2]{Department  of Physics and the Institute for the Early Universe,
 \\
Ewha Womans University, Seoul 120-750, Korea}
\begin{abstract}
Dense neutron matter with recently developed BSk19 and BSk21 Skyrme
effective forces is considered in magnetic fields up to $10^{20}$~G
at zero temperature. The breaking of the rotational symmetry by the
magnetic field leads to the differentiation between the pressures
along and perpendicular to the field direction which becomes
significant in the fields $H>H_{th}\sim10^{18}$~G. The longitudinal
pressure
 vanishes in the critical field
$10^{18}<H_c\lesssim10^{19}$~G, resulting in the longitudinal
instability of neutron matter.  For the Skyrme force fitted to the
stiffer underlying equation of state  (BSk21 vs. BSk19) the
threshold $H_{th}$ and critical $H_c$ magnetic fields become larger.
The longitudinal and transverse pressures as well as
 the anisotropic equation of state of neutron matter are determined under the
conditions relevant for the cores of magnetars.
\end{abstract}


\end{frontmatter}


\section{Introduction}
\label{I}Magnetars are strongly magnetized neutron stars~\cite{DT}
with emissions powered by the dissipation of magnetic energy.
According to one of the conjectures, magnetars can be the source of
the extremely powerful short-duration $\gamma$-ray
bursts~\cite{U,KR,HBS,CK}. The magnetic field strength at the
surface of a magnetar is  of about
$10^{14}$-$10^{15}$~G~\cite{TD,IShS}. Such huge magnetic fields can
be inferred from observations of magnetar periods and spin-down
rates, or from hydrogen spectral lines.  In the interior of a
magnetar the magnetic field strength may be even larger, reaching
values of about $10^{18}$~G~\cite{CBP,BPL}. Under such
circumstances, the issue of interest is the behavior of
 neutron star matter in a strong magnetic
field~\cite{CBP,BPL,CPL,PG,PGNP,PGPR,IY4,IY10}.

In the recent study~\cite{PG}, neutron star matter was approximated
by  pure neutron matter in a model with the effective nuclear
forces. It was shown that the behavior of spin polarization of
neutron matter in the high density region in a strong magnetic field
crucially depends on whether neutron matter develops spontaneous
spin polarization (in the absence of a magnetic field) at  several
times  nuclear matter saturation density, or the appearance of
spontaneous polarization is not allowed  at the relevant densities
(or delayed to much higher densities). The first case  is usual for
the Skyrme forces~\cite{R,S,O,VNB,RPLP,ALP,MNQN,KW94,I,IY,I06},
while the second one is characteristic  for the realistic
nucleon-nucleon (NN) interaction~\cite{PGS,BK,H,VPR,FSS,KS,S11,B}.
In the former case, a ferromagnetic transition to a totally spin
polarized state occurs while in the latter case a ferromagnetic
transition is excluded at all relevant densities and spin
polarization remains quite low even in the high density region.

The scenario for the evolution of spin polarization at high
densities in which the spontaneous ferromagnetic transition in
neutron matter is absent was considered for the magnetic fields up
to $10^{18}$~G~\cite{PG}. However, it was argued in the recent
study~\cite{FIKPS} that in the core of a magnetar the local values
of the magnetic field strength could be as large as $10^{20}$~G, if
to assume the inhomogeneous distribution of the matter density and
magnetic field inside a neutron star, or to allow the formation of a
quark core in the high-density interior of a neutron star
(concerning the last point, see also Ref.~\cite{T}). Under such
circumstances, a different scenario is possible in which a
field-induced ferromagnetic phase transition occurs in the magnetar
core. This idea was explored in the recent research~\cite{BRM},
where it was shown within the framework of a lowest constrained
variational approach with the Argonne $V_{18}$ NN potential that a
fully spin polarized state in neutron matter  could be formed in the
magnetic field $H\gtrsim 10^{19}$~G. Note, however, that, as was
pointed out in the works~\cite{FIKPS,Kh}, in such ultrastrong
magnetic fields the breaking of the ${\cal O}(3)$ rotational
symmetry by the magnetic field results in the
 anisotropy of the total pressure, having a smaller value along than
perpendicular to the field direction. The possible outcome could be
the gravitational collapse of a magnetar along the magnetic field,
if the magnetic field strength is large enough.
  Thus, exploring the possibility of a
field-induced ferromagnetic phase transition in neutron matter in a
strong magnetic field, the effect of the pressure anisotropy has to
be taken into account because this kind of instability could prevent
the formation of a fully polarized state in neutron matter. This
effect was not considered in Ref.~\cite{BRM}, thus, leaving the
possibility of the formation of a fully polarized
 state of neutron spins in a strong magnetic field open. In the
 given
 study, we provide a fully self-consistent calculation of the thermodynamic quantities
 of spin polarized
   neutron matter
 taking into account the appearance
 of the pressure anisotropy in a strong magnetic field. We consider spin polarization
 phenomena in a degenerate magnetized system of
 strongly interacting neutrons within the framework of a Fermi
 liquid  approach~\cite{AKPY,AIP,AIPY,IY3}, unlike to the previous
 works~\cite{FIKPS,Kh}, where interparticle interactions were switched off.

Note that recently new parametrizations of Skyrme forces were
suggested, BSk19-BSk21~\cite{GCP}, aimed to avoid the spontaneous
spin instability of nuclear matter at densities beyond the nuclear
saturation density for vanishing temperature. This is achieved by
adding new density-dependent terms to the standard Skyrme
interaction. The BSk19 parametrization was constrained to reproduce
the equation of state (EoS) of nonpolarized neutron matter~\cite{FP}
obtained in variational calculation with the use of the realistic
Urbana $v_{14}$ nucleon-nucleon potential and  the three-body force
called there TNI. The BSk20 force corresponds to the stiffer
EoS~\cite{APR}, obtained in variational calculation with the use of
the realistic Argonne $V_{18}$ two-body potential and the
semiphenomenological UIX$^*$ three-body force which includes also a
relativistic boost correction $\delta v$. Even a stiffer neutron
matter EoS was suggested in the Brueckner-Hartree-Fock calculation
of Ref.~\cite{LS} based on the same $V_{18}$ two-body potential and
a more  realistic three-body force containing different
meson-exchange contributions. This EoS is the underlying one for the
BSk21 Skyrme interaction.  Further we would like to contrast the
results obtained with the Skyrme forces constrained to
soft~\cite{FP} and most stiff~\cite{LS} underlying EoS, and, by this
reason, choose the BSk19 and BSk21 parametrizations in the
subsequent analysis.

 At
this point, it is worthy to note that  we consider thermodynamic
properties of spin polarized states in neutron  matter in a strong
magnetic field up to the high density region relevant for
astrophysics. Nevertheless, we take into account the nucleon degrees
of freedom only, although other degrees of freedom, such as pions,
hyperons, kaons, or quarks could be important at such high
densities.

\section{Basic equations}  The normal (nonsuperfluid)
states of neutron matter are described
  by the normal distribution function of neutrons $f_{\kappa_1\kappa_2}=\mbox{Tr}\,\varrho
  a^+_{\kappa_2}a_{\kappa_1}$, where
$\kappa\equiv({\bf{p}},\sigma)$, ${\bf p}$ is momentum, $\sigma$ is
the projection of spin on the third axis, and $\varrho$ is the
density matrix of the system~\cite{I,IY,I06}.  The energy of the
system is specified as a functional of the distribution function
$f$, $E=E(f)$, and determines the single particle energy
 \begin{eqnarray}
\varepsilon_{\kappa_1\kappa_2}(f)=\frac{\partial E(f)}{\partial
f_{\kappa_2\kappa_1}}. \label{1} \end{eqnarray} The self-consistent
matrix equation for determining the distribution function $f$
follows from the minimum condition of the thermodynamic
potential~\cite{AKPY,AIP} and is
  \begin{align}\label{2}
 f&=\left\{\mbox{exp}(Y_0\varepsilon+Y_i\cdot \mu_n\sigma_i+
Y_4)+1\right\}^{-1}\\ &\equiv
\left\{\mbox{exp}(Y_0\xi)+1\right\}^{-1}.\nonumber \end{align} Here
the quantities $\varepsilon, Y_i$ and $Y_4$ are matrices in the
space of $\kappa$ variables, with
$\bigl(Y_{i,4}\bigr)_{\kappa_1\kappa_2}=Y_{i,4}\delta_{\kappa_1\kappa_2}$,
$Y_0=1/T$, $Y_i=-H_i/T$ and $ Y_{4}=-\mu_0/T$  being
 the Lagrange multipliers, $\mu_0$ being the chemical
potential of  neutrons, and $T$  the temperature. In Eq.~\p{2},
$\mu_n=-1.9130427(5)\mu_N\approx-6.031\cdot10^{-18}$ MeV/G is the
neutron magnetic moment~\cite{A} ($\mu_N$ being the nuclear
magneton), $\sigma_i$ are the Pauli matrices. Note that, unlike to
Refs.~\cite{IY4,IY10}, the term with the external magnetic field
$\bf H$ is not included in the single particle energy $\varepsilon$
but is separately introduced in the exponent of the Fermi
distribution~\p{2}.

Further it will be assumed that the third axis is directed along the
external magnetic field $\bf{H}$. Given the possibility for
alignment of neutron spins along or opposite to the magnetic field
$\bf H$, the normal distribution function of neutrons and the matrix
quantity  $\xi$ (which we will also call a single particle energy)
can be expanded in the Pauli matrices $\sigma_i$ in spin
space
\begin{align} f({\bf p})&= f_{0}({\bf
p})\sigma_0+f_{3}({\bf p})\sigma_3,\label{7.2}\\
\xi({\bf p})&= \xi_{0}({\bf p})\sigma_0+\xi_{3}({\bf p})\sigma_3.
\end{align}


The distribution functions $f_0,f_3$ satisfy the normalization
conditions
\begin{align} \frac{2}{\cal
V}\sum_{\bf p}f_{0}({\bf p})&=\varrho,\label{3.1}\\
\frac{2}{\cal V}\sum_{\bf p}f_{3}({\bf
p})&=\varrho_\uparrow-\varrho_\downarrow\equiv\Delta\varrho.\label{3.2}
 \end{align}
 Here $\varrho=\varrho_{\uparrow}+\varrho_{\downarrow}$ is the total density of
 neutron matter, $\varrho_{\uparrow}$ and $\varrho_{\downarrow}$  are the neutron number densities
 with spin up and spin down,
 respectively. The
quantity $\Delta\varrho$  may be regarded as the neutron spin order
parameter which  determines the magnetization of the system $M=\mu_n
\Delta\varrho$. The spin ordering of neutrons can also be
characterized by the spin polarization parameter
\begin{align*}
    \Pi=\frac{\Delta\varrho}{\varrho}.
\end{align*}
 The magnetization may
contribute to the internal magnetic field $\tbf{B}=\tbf{H}+4\pi
\tbf{M}$. However, we will assume, analogously to the previous
studies~\cite{BPL,PG,IY4}, that, because of the tiny value of the
neutron magnetic moment, the contribution of the magnetization to
the inner magnetic field $\bf{B}$ remains small for all relevant
densities and magnetic field strengths, and, hence, \bal
\bf{B}\approx \bf{H}.\label{approx}\end{align} Indeed, e.g., the
magnetic field necessary to produce a fully polarized spin state is,
at least, greater than  $10^{19}$~G at the densities relevant for
the cores of magnetars (as will be shown later), while for totally
spin polarized neutron matter with the density
$\varrho=1\,\mbox{fm}^{-3}$ the contribution of the term with the
magnetization to the inner magnetic field amounts at $4\pi M\simeq
1.2\cdot10^{17}$~G.

In order to get the self--consistent equations for the components of
the single particle energy, one has to set the energy functional of
the system. It represents the sum of the matter and field energy
contributions
\begin{equation}\label{en}
E(f,H)=E_m(f)+\frac{H^2}{8\pi}{\cal V}.
\end{equation}
The matter energy is the sum of the kinetic and Fermi-liquid
interaction energy terms~\cite{IY,I06}
\begin{align} E_m(f)&=E_0(f)+E_{int}(f),\label{enfunc} \\
{E}_0(f)&=2\sum\limits_{ \bf p}^{}
\underline{\varepsilon}_{\,0}({\bf p})f_{0}({\bf p}),\nonumber
\\ {E}_{int}(f)&=\sum\limits_{ \bf p}^{}\{
\tilde\varepsilon_{0}({\bf p})f_{0}({\bf p})+
\tilde\varepsilon_{3}({\bf p})f_{3}({\bf p})\},\nonumber \end{align}
where
\begin{align}\tilde\varepsilon_{0}({\bf p})&=\frac{1}{2\cal
V}\sum_{\bf q}U_0^n({\bf k})f_{0}({\bf
q}),\;{\bf k}=\frac{{\bf p}-{\bf q}}{2}, \label{ve0}\\
\tilde\varepsilon_{3}({\bf p})&=\frac{1}{2\cal V}\sum_{\bf
q}U_1^n({\bf k})f_{3}({\bf q}). \label{ve3}
\end{align}
Here  $\underline\varepsilon_{\,0}({\bf p})=\frac{{\bf
p}^{\,2}}{2m_{0}}$ is the free single particle spectrum, $m_0$ is
the bare mass of a neutron, $U_0^n({\bf k}), U_1^n({\bf k})$ are the
normal Fermi liquid (FL) amplitudes, and
$\tilde\varepsilon_{0},\tilde\varepsilon_{3}$ are the FL corrections
to the free single particle spectrum. Taking into account
Eqs.~\p{1},\p{2} and \p{enfunc},  expressions for the components of
the single particle energy read \bal\xi_{0}({\bf
p})&=\underline{\varepsilon}_{\,0}({\bf
p})+\tilde\varepsilon_{0}({\bf p})-\mu_0,\; \xi_{3}({\bf
p})=-\mu_nH+\tilde\varepsilon_{3}({\bf p}).\label{14.2}
\end{align}

In Eqs.~\p{14.2}, the quantities
$\tilde\varepsilon_{0},\tilde\varepsilon_{3}$ are the functionals of
the distribution functions $f_0,f_3$ which, using Eqs.~\p{2} and
\p{7.2}, can be expressed, in turn, through the quantities $\xi$:
\begin{align}
f_{0}&=\frac{1}{2}\{n(\omega_{+})+n(\omega_{-}) \},\label{2.4}
 \\
f_{3}&=\frac{1}{2}\{n(\omega_{+})-n(\omega_{-})\},\label{2.5}
 \end{align} where
 $$
    n(\omega_\pm)=\{\exp(Y_0\omega_\pm)+1\}^{-1},\quad
\omega_{\pm}=\xi_{0}\pm\xi_{3}.$$

The quantity $\omega_{\pm}$, being the exponent in the Fermi
distribution function $n$, plays the role of the quasiparticle
spectrum. The branches $\omega_{\pm}$ correspond to neutrons with
spin up and spin down.

Thus, Eqs.~\p{14.2}--\p{2.5} form the self-consistency equations for
the components of the single particle energy, which should be solved
jointly with the normalization conditions~\p{3.1}, \p{3.2}.

The pressures (longitudinal and transverse with respect to the
direction of the magnetic field) in the system are related to the
diagonal elements of the stress tensor whose explicit expression
reads~\cite{LLP}

\begin{equation}\label{sigma}
    \sigma_{ik}=\biggl[\tilde{ \mathfrak{f}}-\varrho\biggl(\frac{\partial
    \tilde{ \mathfrak{f}}}{\partial \varrho}\biggr)_{{\bf
    H},T}\biggr]\delta_{ik}+\frac{H_iB_k}{4\pi}.
\end{equation}
Here
\begin{equation}\label{Ft}
\tilde{ \mathfrak{f}}=\mathfrak{f}_H-\frac{H^2}{4\pi},
\end{equation}
$\mathfrak{f}_H=\frac{1}{\cal V}(E-TS)-\mathbf{HM}$ is the Helmholtz
free energy density. For the isotropic medium, the stress
tensor~\p{sigma} is symmetric.

The transverse  $p_{t}$  and longitudinal $p_{l}$  pressures are
determined from the formulas
\begin{equation*}
p_{t}=-\sigma_{11}=-\sigma_{22},\; p_{l}=-\sigma_{33}.
\end{equation*}
At zero temperature, using Eqs.~\p{en}, \p{sigma}, one can get the
approximate expressions
\begin{align}\label{press}
p_t&=\varrho\Bigl(\frac{\partial
    e_m}{\partial \varrho}\Bigr)_{H}-e_m+\frac{H^2}{8\pi},\\
    p_l&=\varrho\Bigl(\frac{\partial
    e_m}{\partial \varrho}\Bigr)_{H}-e_m-\frac{H^2}{8\pi},
    \end{align}
where $e_m$ is the  matter energy density, and we disregarded the
higher order small terms containing $M$. The structure of the
pressures $p_t$ and $p_l$ is different that reflects the breaking of
the rotational symmetry by the magnetic field. In ultrastrong
magnetic fields, the quadratic on the magnetic field term (the
Maxwell term) will be dominating, leading to increasing the
transverse pressure and to decreasing the longitudinal pressure.
Hence, at some critical magnetic field, the longitudinal pressure
will vanish, resulting in the longitudinal instability  of neutron
matter. The question then arises: What is the magnitude of the
critical field and
 the corresponding maximum degree of spin polarization in neutron
matter?

\section{Longitudinal and transverse pressures. Anisotropic EoS at zero temperature}
In order to solve the self-consistent equations,  we  utilize the
BSk19 and BSk21 parametrizations of the
 Skyrme interaction, developed in Ref.~\cite{GCP} and generalizing the
 conventional Skyrme parametrizations.  By choosing these Skyrme
 forces, we would like
to study the influence of the underlying EoS, to which these Skyrme
forces were constrained, on thermodynamic quantities of strongly
magnetized dense neutron matter.

The normal FL amplitudes in Eqs.~\p{ve0},\p{ve3} can be related to
the parameters of the Skyrme interaction by formulas~\cite{IY10a}
\bal U_0^n({\bf k})&=2t_0(1-x_0)+\frac{t_3}{3}\varrho^\alpha(1-x_3)
+\frac{2}{\hbar^2}[t_1(1-x_1)\label{101}\\&
+t_4(1-x_4)\varrho^\beta+3t_2(1+x_2)+ 3t_5(1+x_5)\varrho^\gamma]{\bf
k}^{2},
\nonumber\\
U_1^n({\bf
k})&=-2t_0(1-x_0)-\frac{t_3}{3}\varrho^\alpha(1-x_3)+\frac{2}{\hbar^2}[t_2(1+x_2)
\label{102}\\&\quad
+t_5(1+x_5)\varrho^\gamma-t_1(1-x_1)- t_4(1-x_4)\varrho^\beta]{\bf
k}^{2}.\nonumber \end{align}

In these equations, the terms with the  factors $t_4$ and $t_5$ are
the additional density-dependent terms generalizing the $t_1$ and
$t_2$ terms in the conventional form of the Skyrme
interaction~\cite{VB}. They were added  to the usual form with the
aim to avoid the appearance of spontaneous spin instabilities in
nuclear and neutron matter at high densities.  Specific values of
the parameters $t_i, x_i, \alpha, \beta$ and $\gamma$ as well as of
the nuclear saturation density $\varrho_0$ for each parametrization
are given in Ref.~\cite{GCP}.

\begin{figure}[tb]
\begin{center}
\includegraphics[width=8.6cm,keepaspectratio]{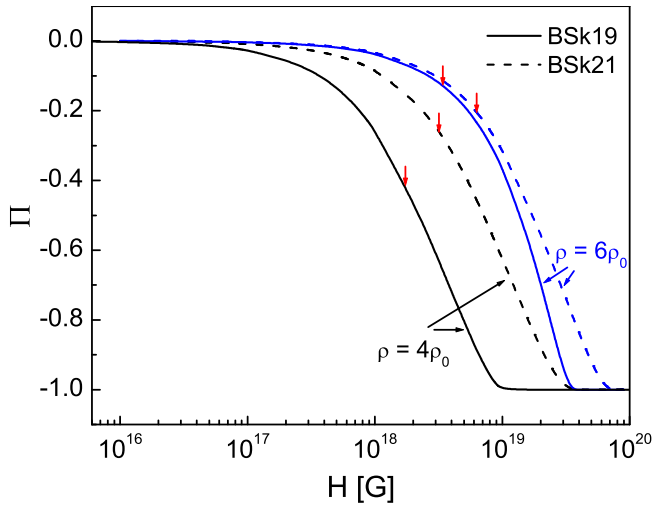}
\end{center}
\vspace{-2ex} \caption{(Color online) Neutron spin polarization
parameter as a function of the magnetic field $H$ for the Skyrme
forces BSk19 (solid lines) and BSk21 (dashed lines) at zero
temperature and fixed values of the density, $\varrho=4\varrho_0$
(two lower branches) and $\varrho=6\varrho_0$ (two upper branches).
The vertical arrows indicate the maximum magnitude of spin
polarization attainable for the corresponding Skyrme force at the
given density, see further details in the text. }
\label{fig1}\vspace{-0ex}
\end{figure}

Now we present the results of the numerical solution  of the
self-consistency equations. Fig.~1 shows the spin polarization
parameter of neutron matter as a function of the magnetic field $H$
at two different values of the neutron matter density,
$\varrho=4\varrho_0$ and $\varrho=6\varrho_0$, which can be relevant
for the central regions of a magnetar. It is seen that the impact of
the magnetic field remains small up to the field strength
$10^{17}$~G. For the BSk21 force (stiff underlying EoS), the
magnitude of the spin polarization parameter is smaller that that
for the BSk19 force  (soft underlying EoS). For both
parametrizations, the larger the density is, the smaller the effect
produced by the magnetic field on spin polarization of neutron
matter.
 At the magnetic
field $H=10^{18}$~G, usually considered as the maximum magnetic
field strength in the core of a magnetar (according to a scalar
virial theorem~\cite{LS91}), the magnitude of the spin polarization
parameter doesn't exceed $25\%$ for the BSk19 force and $8\%$ for
the BSk21 force (for the densities under consideration). However,
the situation changes if the larger magnetic fields are allowable:
With further increasing the magnetic field strength, the magnitude
of the spin polarization parameter increases till it reaches the
limiting value $\Pi=-1$, corresponding to a fully spin polarized
state. For example, this happens at $H\approx 1.3\cdot 10^{19}$~G
for the BSk19 force and at $H\approx 4.3\cdot 10^{19}$~G for the
BSk21 force at $\varrho=4\varrho_0$, i.e., certainly, for magnetic
fields $H\gtrsim10^{19}$~G. Nevertheless, we should check whether
the formation of a fully spin polarized state in a strong magnetic
field is actually possible by calculating the anisotropic pressure
in dense neutron matter. The meaning of the vertical arrows in
Fig.~1 is explained later in the text.

\begin{figure}[tb]
\begin{center}
\includegraphics[width=8.6cm,keepaspectratio]{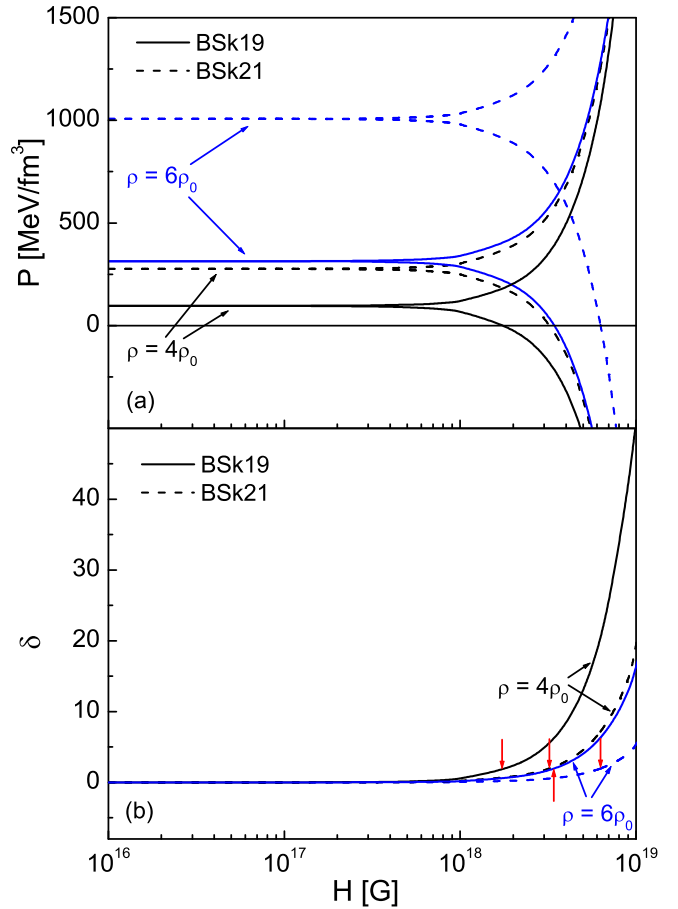}
\end{center}
\vspace{-2ex} \caption{(Color online) (a) Pressures, longitudinal
(descending branches) and transverse (ascending branches),
 as  functions of the magnetic field $H$ for the
Skyrme forces BSk19 (solid lines) and BSk21 (dashed lines) at zero
temperature and  fixed values of the density, $\varrho=4\varrho_0$
(two lower branching curves) and $\varrho=6\varrho_0$ (two upper
branching curves). (b) Same as in the top panel but for the
normalized difference between the transverse and longitudinal
pressures. The vertical arrows in the lower panel indicate the
points corresponding to the onset of the longitudinal instability in
neutron matter. } \label{fig2}\vspace{-0ex}
\end{figure}

Fig.~2a shows the pressures (longitudinal and transverse) in neutron
matter as functions of the magnetic field $H$ at the same densities,
$\varrho=4\varrho_0$ and $\varrho=6\varrho_0$. The upper
 branches in the branching curves correspond to the
transverse pressure, the lower  ones  to the longitudinal pressure.
First, it is clearly seen that up to some threshold  magnetic field
the difference between transverse and longitudinal pressures is
unessential that corresponds to the isotropic regime. Beyond this
threshold magnetic field strength, the anisotropic regime holds for
which the transverse pressure increases with $H$ while the
longitudinal pressure decreases. The stiffer the underlying EoS is
(BSk21 vs. BSk19), the larger the pressure, transverse $p_t$ or
longitudinal $p_l$. Also, the increase of the density has the same
effect on the pressures $p_t$ and $p_l$ as stiffening of the
underlying EoS. The most important feature is that the longitudinal
pressure vanishes at some critical magnetic field $H_c$ marking the
onset of the longitudinal collapse of a neutron star. For example,
$H_c\approx1.7\cdot 10^{18}$~G for BSk19 force and
$H_c\approx3.2\cdot 10^{18}$~G for BSk21 force at
$\varrho=4\varrho_0$, and $H_c\approx3.4\cdot 10^{18}$~G for BSk19
force and $H_c\approx6.3\cdot 10^{18}$~G for BSk21 force at
$\varrho=6\varrho_0$. In all cases under consideration, this
critical value doesn't exceed $10^{19}$~G. In Ref.~\cite{FIKPS}, the
critical field for a relativistic dense gas of free charged fermions
 was found to be close to $10^{19}$~G.

 The magnitude of the spin
polarization parameter $\Pi$ cannot also exceed some limiting value
corresponding to the critical field $H_c$. These maximum values of
the $\Pi$'s magnitude are shown in Fig.~1 by the vertical arrows. In
particular, $\Pi_c\approx-0.42$ for BSk19 force and
$\Pi_c\approx-0.26$ for BSk21 force at $\varrho=4\varrho_0$, and
$\Pi_c\approx-0.13$ for BSk19 force and $\Pi_c\approx-0.20$ for
BSk21 force at $\varrho=6\varrho_0$. As can be inferred from these
values, the appearance of the negative longitudinal pressure  in an
ultrastrong magnetic field prevents the formation of a fully spin
polarized  state in the core of a magnetar. Therefore, only the
onset of a field-induced ferromagnetic phase transition, or its
close vicinity,     can be catched under increasing the magnetic
field strength in dense neutron matter. A complete spin polarization
 in the magnetar
core is not allowed by the appearance of the negative pressure along
the direction of the magnetic field, contrary to the conclusion of
Ref.~\cite{BRM} where the pressure anisotropy in a strong magnetic
field was disregarded.

Fig.~2b shows the difference between the transverse and longitudinal
pressures normalized to the value of the pressure $p_0$ in the
isotropic regime (which corresponds to the weak field limit with
$p_l=p_t=p_0$):
\begin{align*}
    \delta=\frac{p_{t}-p_{l}}{p_0}.
\end{align*}
It is quite reasonable to admit that, when the anisotropic regime
sets in, the splitting between the transverse and longitudinal
pressures becomes comparable with the value of the pressure in the
isotropic regime~\cite{FIKPS}. Applying for the transition from the
isotropic regime to the anisotropic one the approximate criterion
$\delta\simeq 1$, the transition occurs at the threshold field
$H_{th}\approx 1.2\cdot 10^{18}$~G for BSk19 force and
$H_{th}\approx2.2\cdot 10^{18}$~G for BSk21 force at
$\varrho=4\varrho_0$, and at $H_{th}\approx2.3\cdot 10^{18}$~G for
BSk19 force and $H_{th}\approx4.6\cdot 10^{18}$~G for BSk21 force at
$\varrho=6\varrho_0$. In all cases under consideration, the
threshold field $H_{th}$ is larger than $10^{18}$~G, and, hence, the
isotropic regime holds for the fields up to $10^{18}$~G. For
comparison, the threshold field for a relativistic dense gas of free
charged fermions
 was found to be about $10^{17}$~G~\cite{FIKPS} (without including
 the anomalous magnetic moments of fermions). For a degenerate
 gas of free neutrons the model dependent estimate
 gives $H_{th}\simeq4.5\cdot 10^{18}$~G~\cite{Kh} (including
 the neutron anomalous magnetic moment).
The normalized splitting of the transverse and longitudinal
pressures increases more rapidly with the magnetic field at the
smaller density and/or for the Skyrme force BSk19 with the softer
underlying EoS.   The vertical arrows in Fig.~2b indicate the points
corresponding to the onset of the longitudinal instability in
neutron matter. Since the threshold field $H_{th}$ is less than the
critical field $H_c$ for the appearance of the longitudinal
instability, the anisotropic regime can be relevant for the core of
a magnetar.  The maximum allowable normalized splitting of the
pressures corresponding to the critical field $H_c$ is $\delta\sim
2$. If the anisotropic regime sets in, a neutron star
has the  oblate form. 
Thus, as follows from the preceding discussions, in the anisotropic
regime  the pressure anisotropy plays an important role in
determining the spin structure and configuration of a neutron star.

\begin{figure}[tb]
\begin{center}
\includegraphics[width=8.6cm,keepaspectratio]{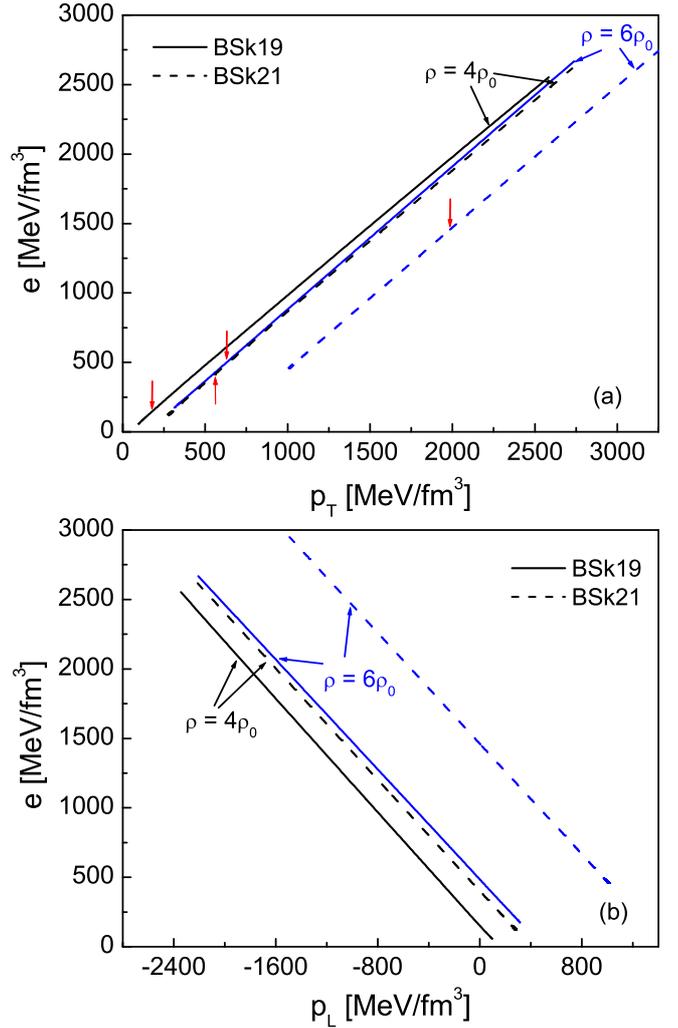}
\end{center}
\vspace{-2ex} \caption{(Color online) The energy density of the
system  as  a function of: (a) the transverse pressure $p_t$, (b)
the longitudinal  pressure $p_l$  for the Skyrme forces BSk19 (solid
lines) and BSk21 (dashed lines) at zero temperature and fixed values
of the density, $\varrho=4\varrho_0$ and $\varrho=6\varrho_0$. The
meaning of the  vertical arrows in the top panel is the same as in
Fig.~2.  In the bottom panel, the physical region corresponds to
$p_l>0$.} \label{fig3}\vspace{-0ex}
\end{figure}

Because of the pressure anisotropy, the EoS of neutron matter in a
strong magnetic field is also anisotropic. Fig.~3 shows the
dependence of the energy density of the system on the transverse
pressure (top panel) and on the longitudinal pressure (bottom panel)
at the same densities considered above.
 Since in an ultrastrong magnetic field
 the dominant Maxwell  term enters the pressure $p_t$ and energy
density with  positive sign and the pressure $p_l$ with  negative
sign, the energy density is the increasing function of $p_t$ and
decreasing function of $p_l$. In the case of $e(p_t)$ dependence, at
the given density, the same $p_t$ corresponds to the larger magnetic
field $H$ for the BSk19 force compared with the BSk21 force (see
Fig.~2a). The overall effect of two factors (the stiffness/softness
of the underlying EoS and magnetic field) will be the larger value
of the energy density at the given $p_t$ and density for the BSk19
force compared with the BSk21 force (see Fig.~3a). The analogous
arguments show that, for the given Skyrme parametrization and at the
given $p_t$, the energy density is larger for the smaller density.
In the case of $e(p_l)$ dependence, at the given density, the same
$p_l$ corresponds to the smaller magnetic field $H$ for the BSk19
force compared with the BSk21 force (see Fig.~2a). Hence, the energy
density at the given $p_l$ and density is larger for the BSk21 force
than that for the BSk19 force (see Fig.~3b). Analogously, for the
given Skyrme parametrization and at the given $p_l$, the energy
density is larger for the larger density. In the bottom panel, the
physical region corresponds to the positive values of the
longitudinal pressure. Note that, because of the validity of the
approximation given by Eq.~\p{approx}, the energy density containing
the field energy contribution is practically indistinguishable from
the Helmholtz free energy density.

It is worthy to notice that the occurrence of the longitudinal
instability in a strong magnetic field will lead to the compression
of a neutron star along the magnetic field with a subsequent
increase of the density.
 Such an increase can eventually cause the appearance of new
particle species. Already for the conditions of high density
considered in this study one can assume that a deconfined phase of
quarks exists in the interior of a neutron star. Then  a hybrid star
consisting of deconfined quark matter in the core and nuclear matter
in the outer layers can be regarded as a relevant astrophysical
object. As follows from the general arguments presented in our
study, the critical magnetic field strength at which the
longitudinal pressure vanishes inevitably exists for a hybrid star
as well. The determination of the corresponding critical value needs
a separate investigation. As a possible guess for this value, one
can refer to the results of the study of quark matter within the MIT
bag model~\cite{FIKPS}, where it was estimated to be close to
$10^{19}$ G (with the  bag constant put to zero in the final
expressions). Also, the mass-radius relationship is a relevant
characteristic of neutron stars. Usually it is found by solving the
Tolman-Oppenheimer-Volkoff (TOV) equations~\cite{TOV} for a
spherically symmetric and static neutron star. In an ultrastrong
magnetic field, the EoS becomes essentially anisotropic. Unlike to
the standard scheme,
  the mass-radius relationship should be found
by the self-consistent treatment of the anisotropic EoS and
axisymmetric TOV equations substituting the conventional TOV
equations in the case of  an axisymmetric neutron star.

Note that in this research we have studied the impact of a strong
magnetic field on thermodynamic properties of dense neutron matter
at zero temperature. It would be also of interest to extend this
research to finite temperatures relevant for proto-neutron stars
which can lead to a number of interesting effects, such as, e.g., an
unusual behavior of the entropy of a spin polarized
state~\cite{IY2,I07}.

In summary, we have considered spin polarized states in dense
neutron matter in the model with the Skyrme effective NN interaction
(BSk19 and BSk21 parametrizations) under the presence of strong
magnetic fields up to $10^{20}$~G. It has been shown that in the
magnetic field $H>H_{th}\sim10^{18}$~G the pressure anisotropy has a
significant impact on thermodynamic properties of neutron matter. In
particular, vanishing of the pressure along the direction of the
magnetic field in the critical field $H_c>H_{th}$ leads to the
appearance of the longitudinal instability  in neutron matter. For
the Skyrme force with the stiffer underlying EoS  (BSk21 vs. BSk19),
  the threshold $H_{th}$ and critical $H_c$ magnetic
fields become larger. The increase of the density of neutron matter
also leads to increasing the fields $H_{th}$ and $H_c$. Even in the
extreme scenario with the most stiff underlying EoS considered in
this work and at the densities about  several times nuclear
saturation density, the critical field $H_c$ doesn't exceed
$10^{19}$~G which can be considered as the upper bound on the
magnetic field strength inside a magnetar. Our calculations show
that the appearance of the longitudinal instability prevents the
formation of  a fully spin polarized state in neutron matter, and
only the states with mild spin polarization can be developed.  The
longitudinal and transverse pressures and anisotropic EoS of neutron
matter in a strong magnetic field have been determined at the
densities relevant for the cores of magnetars. The obtained results
can be of importance in the structure studies of strongly magnetized
neutron stars.

J.Y. was supported by grant 2010-0011378 from Basic Science Research
Program through NRF of Korea funded by MEST and by  grant R32-10130
from WCU project of MEST and NRF.








\end{document}